\documentclass[conference]{IEEEtran}
\IEEEoverridecommandlockouts
\usepackage{cite}
\usepackage{amsmath,amssymb,amsfonts}
\usepackage{algorithmic}
\usepackage{graphicx}
\usepackage{textcomp}
\usepackage{xcolor}
\usepackage{hyperref}
\usepackage{cleveref}
\def\BibTeX{{\rm B\kern-.05em{\sc i\kern-.025em b}\kern-.08em
    T\kern-.1667em\lower.7ex\hbox{E}\kern-.125emX}}

\DeclareMathOperator*{\argmin}{arg\,min}
\begin{document}

\title{iMagLS: Interaural Level Difference with Magnitude Least-Squares Loss for Optimized First-Order Head-Related Transfer Function}


\author{\IEEEauthorblockN{Or Berebi\IEEEauthorrefmark{1},
Zamir Ben-Hur\IEEEauthorrefmark{2},
David Lou Alon\IEEEauthorrefmark{2} and
Boaz Rafaely\IEEEauthorrefmark{1}}
\IEEEauthorblockA{\IEEEauthorrefmark{1}School of Electrical and Computer Engineering, Ben-Gurion University of the Negev,
Beer-Sheva 84105, Israel}
\IEEEauthorblockA{\IEEEauthorrefmark{2}Reality Labs Research, Meta, 1 Hacker Way, Menlo Park, CA 94025, USA}}

\maketitle

\begin{abstract}
Binaural reproduction for headphone-based listening is an active research area due to its widespread use in evolving technologies such as augmented and virtual reality (AR and VR). On the one hand, these applications demand high quality spatial audio perception to preserve the sense of immersion. On the other hand, recording devices may only have a few microphones, leading to low-order representations such as first-order Ambisonics (FOA). However, first-order Ambisonics leads to limited externalization and spatial resolution. In this paper, a novel head-related transfer function (HRTF) preprocessing optimization loss is proposed, and is minimized using nonlinear programming. The new method, denoted iMagLS, involves the introduction of an interaural level difference (ILD) error term to the now widely used MagLS optimization loss for the lateral plane angles. Results indicate that the ILD error could be substantially reduced, while the HRTF magnitude error remains similar to that obtained with MagLS. These results could prove beneficial to the overall spatial quality of first-order Ambisonics, while other reproduction methods could also benefit from considering this modified loss.
\end{abstract}


\section{Introduction}
In recent years, binaural sound reproduction has gained significant attention due to its ability to create an immersive and realistic listening experience. Binaural sound refers to the technique of capturing and reproducing sound in a way that mimics natural listening when using headphones. One format that has gained popularity and seen widespread adoption for binaural sound reproduction is Ambisonics\cite{gerzon1973periphony}. Ambisonics signals are typically derived from recordings with spherical microphone arrays and filtered with measured or modeled head-related transfer function HRTF to produce binaural signals~\cite{zotter2019ambisonics,avni2013spatial}.

First-order Ambisonics (FOA) is a commonly used format in Ambisonics due to the relatively simple microphone array required for recording (only four microphones), established theoretical and algorithmic literature, and availability and broad adoption in applications such as $360^\circ$ video, surround sound, and virtual reality (VR)~\cite{zotter2019ambisonics}. However, FOA has shortcomings with regard to spatial resolution and timbre degradation that limit its application for VR audio~\cite{rafaely2022spatial}.

To improve the listening experience for signals processed with FOA, researchers have proposed various HRTF preprocessing methods. These methods aim to overcome some of the spatial resolution and timbre degradation issues of FOA by correcting the low-order HRTF spatial and spectral errors. Examples of these methods include global equalization, time alignment, ear alignment, and magnitude least squares (MagLS)~\cite{ben2017spectral, zaunschirm2018binaural,ben2019efficient, schorkhuber2018binaural}.

While the MagLS method has been proven to be very beneficial in reducing spectral error in FOA, the resulting binaural signals still have significant spatial errors~\cite{zotter2019ambisonics, rafaely2022spatial}. The aim of this paper is to propose and investigate improvements over the current MagLS method. This is achieved by the use of the HRTF preprocessing with integrated MagLS and interaural level difference (ILD) optimized errors. This approach aims to improve FOA's spatial attributes while preserving low spectral errors, ultimately providing a more immersive and realistic audio experience for listeners.

\section{Problem Formulation}
In Ambisonics, binaural signals can be rendered in the spherical harmonics (SH) domain by combining the Ambisonics signal with the left-ear and right-ear HRTFs~\cite{xie2013head}. This can be expressed as~\cite{rafaely2010interaural}:
\begin{equation}\label{eq:Ambsionics_reproduction}
p^{L / R}(f) = \sum_{n=0}^{N} \sum_{m=-n}^{n} [\tilde{a}_{nm}(f)]^* h_{nm}^{L / R}(f),
\end{equation}
where $p^{L / R}(f)$ denotes the binaural signal for the left/right ear and $f$ denotes the frequency. The $N$'th order Ambisonics signal is denoted as $a_{nm}(f)$ and is modified to $\tilde{a}_{nm}(f)= (-1)^m[a_{n(-m)}(f)]^*$, encoding the sound-field information that can be captured using a spherical microphone array~\cite{zotter2019ambisonics}. Henceforth we refer only to the left-ear for simplicity, while the same operations are conducted for the right ear. The $(N+1)^2$ left-ear HRTF SH coefficients are represented by $h^L_{nm}(f)$, which can be obtained by minimizing the following generalized problem:
\begin{align}\label{eq:find_w_hat_continuous}
&h^{L}_{nm}(f) = \\ 
&\argmin_{\hat{h}^{L}_{nm}(f)} \int_{\Omega \in \mathcal{S}^2} D \left( \sum_{n=0}^{N} \sum_{m=-n}^{n} \hat{h}^{L}_{nm}(f) Y_n^m(\Omega), H(f,\Omega) \right) d\Omega, \nonumber
\end{align}
with $H(f,\Omega)$ corresponding to a reference measured or modeled HRTF, and the spherical harmonic function is represented by $Y_n^m(\Omega)$ with degree $m$ and order $n$. The complex coefficients $h^{L}_{nm}(f)$ serve to minimize the distance function $D(.,.)$ across a collection of directions $\Omega=(\varphi,\theta)\in \mathcal{S}^2$, with $D(.,.)$ representing a measure of dissimilarity.

The choice of function $D(.,.)$ can greatly impact the spectral and spatial quality of $p^L(f)$.  For example, using $D(x,y) = |x - y |^2$, the least squares (LS) distance yields a closed-form solution to Eq.\ref{eq:find_w_hat_continuous} given by the inverse spherical harmonic transform (ISHT) of $H(f,\Omega)$\cite{rafaely2015fundamentals}. However, when $N$ is small, particularly with $N=1$, the LS solution results in a $p^L(f)$ that is not perceptually comparable to its high-order counterpart, with high-frequency rolloff and limited spatial resolution\cite{zotter2019ambisonics}. Another example is using $D(x,y) = |\, |x| - |y| \, |^2$, the magnitude least squares (MagLS) distance. Solving Eq.\ref{eq:find_w_hat_continuous} for the MagLS distance for frequencies greater than a cutoff frequency $f_c$ greatly improves the spectral quality of $p^L(f)$ compared to the LS method even for low $N$, as shown in~\cite{schorkhuber2018binaural}. Unfortunately, the spatial attributes of the MagLS solution are still poor when considering $N=1$ reproduction~\cite{zotter2019ambisonics,rafaely2022spatial}.

\section{Proposed Method}
The MagLS formulation has been shown to be effective in addressing rolloff issues, even with $N=1$. However, it is important to acknowledge that the spatial resolution at $N=1$ is still a concern~\cite{zotter2019ambisonics}. To tackle this limitation, we propose a new $D(.,.)$ function that addresses the issue of binaural information while preserving the spectral benefits of the MagLS formulation. This function will be used in the problem formulation that we present in the following section. Note that for the rest of this paper, we omit the superscript $L$ for abbreviation.

The ILD is known to be an important binaural cue for sound localization~\cite{blauert1997spatial}, and preserving ILD in an FOA HRTF could potentially enhance spatial perception. Consider the following order-$N$ HRTF representation:
\begin{equation}
    \mathcal{SFT}\left( h_{nm}(f) \right) \equiv \sum_{n=0}^{N} \sum_{m=-n}^{n} h_{nm}(f)Y_n^m(\Omega).
\end{equation}
We begin by defining the ILD for a given HRTF as~\cite{xie2013head}:
\begin{equation}\label{eq:ILD}
    \text{ILD}(\Omega_0,f_0) = 10 \log_{10}\frac{\int_{f_1}^{f_2} G(f_0,f)|p^L(\Omega_0,f)|^2 df}{\int_{f_1}^{f_2} G(f_0,f)|p^R(\Omega_0,f)|^2 df} .
\end{equation}
Here, $G(f_0,f)$ denotes the Gammatone function centered at $f_0$, and $p^{L,R}(\Omega_0,f)$ represents the left- and right-ear signals resulting from a single plane wave sound field at an incident angle $\Omega_0$. The ILD is evaluated over the horizontal plane $\Omega_0 \in (\theta = 90^\circ, 0^\circ \le \phi < 360^\circ)$ for frequencies $f_1 \le f \le f_2$. This computation facilitates perceptually-motivated smoothing of the ILD across frequencies. The ILD error as a function of both frequency and direction is defined as follows:
\begin{equation}\label{eq:ILD_error}
    \epsilon_{\text{ILD}}(\Omega_0,f_0) = | \text{ILD}_{ref}(\Omega_0, f_0) - \text{ILD}(\Omega_0, f_0) | .
\end{equation}
Here, $\text{ILD}_{ref}(\Omega_0, f_0)$ and $\text{ILD}(\Omega_0, f_0)$ refer to the ILD of the reference HRTF $H(\Omega_0,f)$ and the order $N$ HRTF $\mathcal{SFT}\left( h_{nm}(f) \right)$, respectively.

Next, the magnitude error between $H(\Omega,f)$ and $\mathcal{SFT}\left( h_{nm}(f) \right)$ is defined in a similar manner to the MagLS formulation:
\begin{equation}\label{eq:mag_error}
    \epsilon_{mag}(\Omega,f) =  | \, |H(\Omega,f)| - |\mathcal{SFT}\left( h_{nm}(f) \right)|  \, |^2 .
\end{equation}
Finally, combining both error terms leads to the following optimization problem:
\begin{equation}\label{eq:iMagLS_optimization}
    h_{nm}(f)= \argmin_{\hat{h}_{nm}(f) } \left[ \sum_{\Omega} \epsilon_{mag}(\Omega) + \lambda \sum_{\Omega_0} \epsilon_{\text{ILD}}(\Omega_0) \right] ,
\end{equation}
with $\epsilon_{mag}(\Omega)$ and $ \epsilon_{\text{ILD}}(\Omega_0)$ both referring to error frequency averaging, and $\lambda \in \mathcal{R} $ is used as a regularization term. The formulation in Eq.\ref{eq:iMagLS_optimization} is referred to as ILD-MagLS (iMagLS for short).

\begin{figure}
 \centerline{\framebox{
 \includegraphics[width=7.6cm]{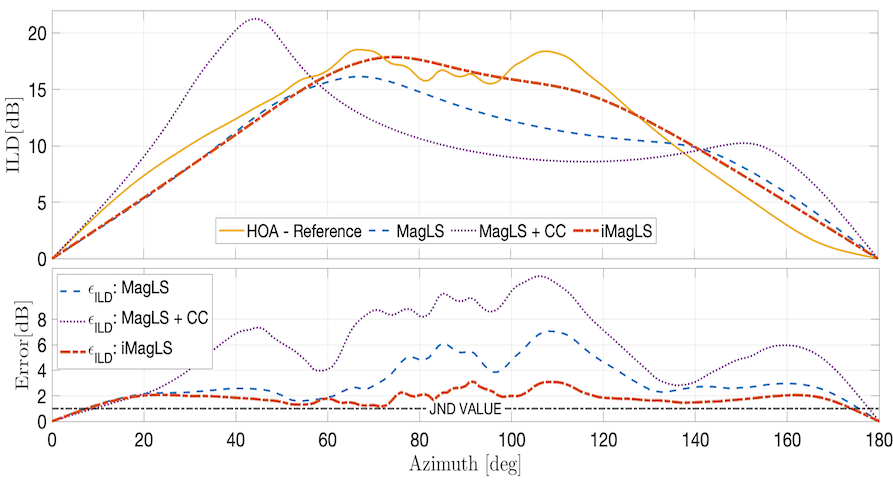}}}
 \caption{ILD curves (top) and ILD error curves (bottom) as a function of $\Omega_0$, averaged over frequency.}
 \label{fig:ILD}
\end{figure}

\section{Numerical Evaluation}
The Broyden–Fletcher–Goldfarb–Shannon Quasi-Newton algorithm~\cite{broyden1970convergence} was used to minimize Eq.\ref{eq:iMagLS_optimization} over $1.2 \le f < 20$ kHz. Note that $\epsilon_{mag}(\Omega)$ and $\epsilon_{ILD}(\Omega_0)$ are both averaged over frequencies. Therefore, the numerical solver minimizes the average error over all frequencies, rather than individually for each frequency. It was observed that this leads to smoother error curves, which may be more perceptually favorable in terms of noticeable spectral artifacts and noise.

The simulated KEMAR HRTF~\cite{burkhard1975anthropometric} was used to evaluate the proposed method, with order $N=35$ as the reference and $N=1$ used to evaluate the MagLS and iMagLS solutions. The evaluation used a Lebedev sampling scheme of order $35$ with $1730$ nearly-uniformly-distributed directions to evaluate $\epsilon_{mag}(\Omega)$. As suggested in~\cite{zotter2019ambisonics}, the MagLS solution presented in this analysis also includes the covariance constraint (MagLS+CC) global EQ variant~\cite{vilkamo2013optimized}. The MagLS solution served as the initial solution for the solver, and $\lambda$ was chosen such that the error terms were equal at the first iteration. The results were evaluated in terms of the magnitude error $\epsilon_{mag}(f)$ averaged over all $\Omega$ directions and in terms of the ILD error $\epsilon_{ILD}(\Omega_0)$ averaged over $1.2 \le f < 20$ kHz.

\begin{figure}
 \centerline{\framebox{
 \includegraphics[width=7.8cm]{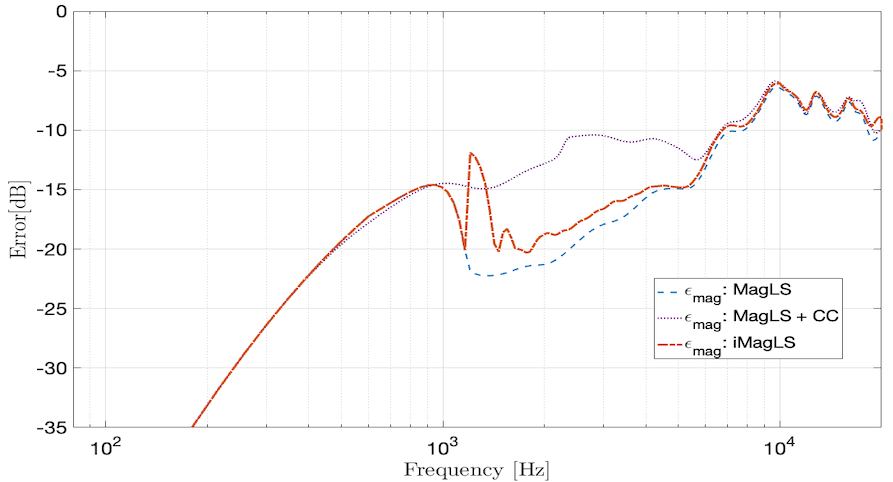}}}
 \caption{Magnitude error as a function of frequency, averaged over $\Omega$ directions.}
 \label{fig:MAG}
\end{figure}

The ILD evaluation is presented in Fig.\ref{fig:ILD}. The ILD of Eq.\ref{eq:ILD} was averaged over frequency; the yellow solid curve represents the KEMAR reference, the dashed blue curve represents the ILD of an $N=1$ MagLS solution, the purple dotted curve represents the ILD of an $N=1$ MagLS+CC solution, and the red dash-dotted curve represents the proposed iMagLS of a similar order. The graph below represents the ILD error of Eq.\ref{eq:ILD_error} averaged over frequency for MagLS (dash blue), MagLS+CC (dotted purple) and iMagLS (dash-dotted red). As shown in Fig.\ref{fig:ILD}, the improvement in terms of ILD is noticeable both in terms of averaged ILD and averaged ILD error. An ILD error of below $2$ dB was preserved by iMagLS for most incident angles, which is close to the Just Notable Differences (JND) ($\sim$1,dB~\cite{mills1960lateralization,yost1988discrimination}), while the MagLS and MagLS+CC error was significantly higher than the JND value for most angles.

The magnitude error of~Eq.\ref{eq:mag_error} averaged over the $\Omega$ directions is presented in Fig.\ref{fig:MAG}. 
The dash blue, dotted purple, and dash-dotted red curves represents MagLS, MagLS+CC, iMagLS error, respectively. Although the iMagLS error was higher for all $f > 1200$ Hz,compared to the optimal MagLS solution,  it was only slightly higher with an averaged absolute difference of $1.72$ dB over these frequencies.

The results suggest that improving the ILD error while maintaining a relatively low binaural magnitude error is possible. We argue that when searching for a low-order $h_{nm}$, one should consider the relation between both ears in addition to the magnitude accuracy of each ear individually. While the rational of the MagLS+CC variant align with this argument, results show that iMagLS outperformed MagLS+CC both in terms of ILD and magnitude error.

\section{Conclusions}
This paper proposed a novel optimization method, iMagLS, to address the limitations of FOA by improving its spatial information shortcomings. The method optimizes HRTF preprocessing with an ILD error term incorporated into the widely used MagLS method. The results show that the proposed method can significantly reduce ILD errors, while maintaining similar HRTF magnitude errors compared to the MagLS method. Based on these results, we propose that selecting a low-order $h_{nm}$ should consider not only the individual magnitude accuracy of each ear but also the interaural relationship between them. Additionally, a more comprehensive study involving a broader range of HRTFs and listening tests is suggested for future work to validate these claims.

\bibliographystyle{IEEEtran}
\bibliography{fa2023_template}

\end{document}